\documentclass[fleqn,usenatbib]{mnras}

\newcommand{\g}{$\gamma$-ray}
\newcommand{\lat}{\textit{Fermi}-LAT}

\usepackage{newtxtext,newtxmath}
\usepackage{soul}

\usepackage[T1]{fontenc}

\DeclareRobustCommand{\VAN}[3]{#2}
\let\VANthebibliography\thebibliography
\def\thebibliography{\DeclareRobustCommand{\VAN}[3]{##3}\VANthebibliography}

\usepackage{graphicx}	
\usepackage{amsmath}	

\title[Berkeley 59]{Diffuse gamma-ray emissions around the stellar cluster Berkeley 59}

\author[Ou, Yang \& Pei]{
Ziwei Ou,$^{1}$\thanks{E-mail: ziwei@sjtu.edu.cn (Ziwei Ou)}
Xiaolong Yang,$^{2,3}$\thanks{E-mail: yangxl@shao.ac.cn (Xiaolong Yang)}
Songpeng Pei,$^{4}$
\\
$^{1}$Tsung-Dao Lee Institute, Shanghai Jiao Tong University, Shanghai, 200240, China\\
$^{2}$Shanghai Astronomical Observatory, Chinese Academy of Sciences, Shanghai 200030, China\\
$^{3}$Shanghai Key Laboratory of Space Navigation and Positioning Techniques, Shanghai 200030, China\\
$^{4}$School of Physics and Electrical Engineering, Liupanshui Normal University, Liupanshui 553004, China
}

\date{Accepted XXX. Received YYY; in original form ZZZ}

\pubyear{\the\year{}}

\begin{document}
\label{firstpage}
\pagerange{\pageref{firstpage}--\pageref{lastpage}}
\maketitle

\begin{abstract}
We report a detailed analysis on the young stellar cluster Berkeley 59 using \textit{Fermi}-LAT. Using up-to-date source catalog and background models, we found significant extended GeV emission around Berkeley 59, which can be modeled by a radial Gaussian of $1.02^{\circ}$ radius with a significance of the extension of 10.6$\sigma$. We investigated the molecular, neutral and ionized gas content and the hadronic origin. The gamma-ray spectrum of Berkeley 59 has a photon index of $2.88 \pm 0.06$. The derived gas mass from H$_2$, HI and HII around Berkeley 59 is $\sim 6.34 \times 10^4 M_{\odot}$. We derived the relationship between cosmic ray acceleration efficiency and diffusion coefficient. Our results suggest that the extended gamma-ray emission originates from cosmic rays accelerated by cluster winds interacting with surrounding gas.
\end{abstract}


\section{Introduction} \label{sec:intro}

The origin of galactic cosmic rays (CRs) remains unknown. For a long time, it has been believed that the acceleration sites of Galactic CRs are supernova remnants (SNRs) \citep{Drury1994,Hillas2005,Uchiyama2007,Gabici2009}. Diffusive shock acceleration is considered as principal CR production mechanism for SNRs \citep{Celli2019,Mitchell2021}. However, growing evidence suggests that stellar clusters are also the excellent energy sources of CRs in the Galaxy \citep{Aharonian2019,Morlino2021,Blasi2023}. Most massive stars form in associations and/or clusters, which serve as important accelerators of very-high-energy (VHE) and ultra-high-energy (UHE) CRs \citep{Gupta2018,Lemoine-Goumard2025}. The accelerated CRs interacting with the surrounding ambient gas may produce the observed \g\ emissions. Nevertheless, only a few stellar clusters have been identified as VHE or UHE \g\ sources, such as Westerlund 1 \citep{Aharonian2022}, Westerlund 2 \citep{Yang2018} and NGC 3603 \citep{Yang2017}. Stellar clusters hosting massive stars may experience numerous supernova explosions that can act as VHE CR accelerators \citep{Shara2002,Rogers2013,Bykov2015,Peron2024a}. A substantial fraction of CRs should be accelerated in young massive stellar clusters containing lots of OB stars and the superbubbles created by these clusters \citep{Ackermann2011,Menchiari2024}. Superbubbles possess sufficient energy originating from stellar winds and supernova explosions \citep{Meyer2024,Sushch2025}. Recent advancements in multi-wavelength observations and simulations have revealed that stellar clusters, especially their stellar wind-driven superbubbles, may contribute significantly to the CR budget \citep{Zhao2024,Harer2025}. Observations of stellar clusters allow us to constrain the fraction of stellar wind energy transferred into \g\ .

Stellar clusters serve as unique laboratories for understanding the properties of the interstellar medium (ISM) and the nature of interactions between young stars and the ISM \citep{Elmegreen2010,Asahina2017,Paron2021}. As such, they provide an alternative testing ground for CR origin theories, potentially helping to resolve discrepancies in the SNR-based paradigm, particularly concerning the maximum particle energy. Studies of the Cygnus cocoon and Westerlund 1 highlight the challenges in distinguishing between leptonic and hadronic origins of high-energy emissions \citep{Vieu2023,Blasi2023}. The commonly invoked 1/r radial profile, often derived from diffusion models that omit key physical processes such as CR advection, distributed acceleration sites, and radiative losses, may not accurately represent the true CR spatial distribution \citep{Yang2018,Yang2022}. Alternative scenarios involving CR injection at wind termination shocks or supernova events have been shown to reproduce the 1/r profile equivalently. Despite these degeneracies, the increasing number of \g\ detections coincident with stellar clusters and the emerging spectral and morphological characterizations of these sources provide circumstantial evidence supporting a significant contribution to the high-energy CR population, although this remains to be conclusively validated \citep{Ge2022,Ge2024,Peron2025b}.

CRs are believed to be important regulators of the star-formation process \citep{Acciari2009,Meijerink2011}. As they travel through the ISM, CRs interact with the ambient gas and radiation fields to produce \g\, which trace the CR population. Therefore, \g\ emissions are among the best tools for studying CR properties in star-forming environments \citep{Li2025}. By analyzing \g\ emission in these systems, we can measure the fraction of interacting high-energy particles as a function of the star formation rate. Efficient particle acceleration makes star-forming regions significant \g\ sources. Star-forming regions, such as G25.0+0.0 \citep{Katsuta2017}, W40 \citep{Sun2020}, Masgomas-6a \citep{Wang2022}, Carina Nebula Complex \citep{Ge2022}, G305 \citep{Liu2024} and W3 \citep{Wu2024}, have been observed by \lat\ to exhibit diffuse \g\ emissions. Massive stars undergo supernova explosions at the end of their lives, enriching the ISM with heavy elements. SNRs can ionize the surrounding medium and contribute a significant pressure component to the ISM which amplifies magnetic fields \citep{Nava2019,Recchia2022}. CRs escape from their acceleration sites, and low-energy CRs are especially dominant sources of ionization in the Galaxy. They penetrate deeply into the dense cores of molecular clouds \citep{Yang2023}.

Berkeley 59 (R.A.=0.570$^{\circ}$, Dec.=67.441$^{\circ}$) is a young ($\sim$ 2 Myr) and nearby ($\sim$ 1 kpc) stellar cluster. It is located in the central region of the OB association Cepheus OB4 surrounded by an H II region \citep{Lata2011,Eswaraiah2012}. The younger pre-main-sequence (PMS) stars tend to concentrate closer to the inner region than to the outer region of the cluster, which is linked to the time evolution of star-forming clouds \citep{Panwar2018}. Considering the central part of the radius $\sim$ 4 arcmin ($\sim$ 1.2 pc), the lower-limit of total mass was estimated to be $\sim 10^3 M_{\odot}$, which is similar to the well-studied young massive stellar clusters (YMSCs) Westerlund 1 and Westerlund 2. Moreover, \cite{Peron2024b} suggests that Berkeley 59 has correlation with unassociated \lat\ sources. Therefore, it serves as a typical yet unique laboratory for studying CR acceleration in young stellar clusters due to its proximity, youth, and multiwavelength detectability.

Section \ref{sec:fermi-data} provides \g\ data analysis with \lat\ data. In Section \ref{sec:gas}, we study gas distribution around Berkeley 59. The CR injection from Berkeley 59 is discussed in Section \ref{sec:injection}. Finally, we summarize our main conclusion in Section \ref{sec:conclu}.
 
\section{\lat\ Analysis} \label{sec:fermi-data}

We selected all events within a $15^{\circ} \times 15^{\circ}$ region of interest (ROI) for 4FGL~J0002.1$+$6721 (R.A.=0.5419$^{\circ}$, Dec.=67.3578$^{\circ}$), which is the closest \lat\ source to Berkeley 59 (offset $4.8'$). The Pass 8 data collected from 4 August 2008 to 30 October 2024 were used to study the GeV emission. The analysis was performed using Fermipy \textit{v1.2.0} and the Fermi Science Tools \textit{v11r5p3} packages. We adopted the Galactic diffuse emission model gll\_iem\_v07 and isotropic extragalactic emission model P8\_R3\_V3 as diffuse background components. The (evclass == 128) and event type FRONT + BACK (evtype == 3) were used in our study. We considered only \g\ events in the 300 MeV – 300 GeV, applying the standard data quality selection criteria “(DATA\_QUAL > 0)\&\&(LAT\_CONFIG==1).” The latest \lat\ source catalog (4FGL-DR4) was used to construct a source model \citep{Ballet2023}. To minimize contamination from \g\ from the Earth's limb, the maximum zenith angle is of $90^{\circ}$ was adopted. To derive the emission related to Berkeley 59, we performed a binned likelihood analysis by using the tool \textit{gtlike}. In the analysis, we include all the sources in the 4FGL-DR4 catalog and the Fermi diffuse background model gll\_iem\_v07.fits for the Galactic \g\ emission, as well as the isotropic background model iso\_P8R3\_SOURCE\_V3\_v1.txt \footnote{https://fermi.gsfc.nasa.gov/ssc/data/access/lat/BackgroundModels.html}. The spectral parameters of sources within $5^{\circ}$ were set free.

\subsection{Spatial Analysis}

\begin{figure}
    \centering
    \includegraphics[width=0.95\linewidth]{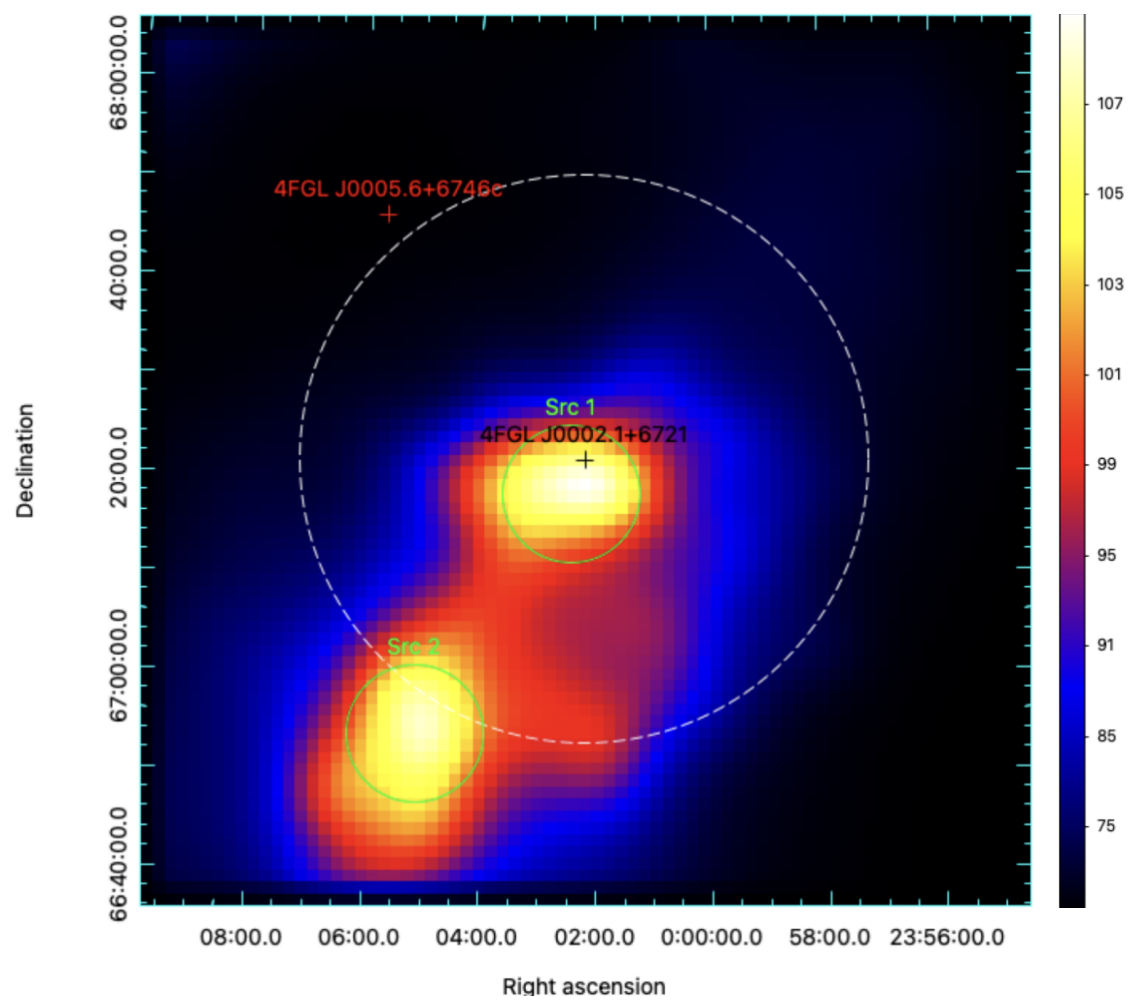}
    \caption{\lat\ TS map between 300 MeV and 300 GeV. The dashed circle shows the Gaussian model with a radius of $1.02^{\circ}$. Src1 and Src2 are shown also.}
    \label{fig:ts-map}
\end{figure}

The test statistic (TS) is adopted to estimate the significance of \g\ sources. It is defined by TS = 2 (ln $L_1$ - ln $L_0$), where $L_1$ and $L_0$ represent maximum likelihood values for background with target source and without target source. We removed 4FGL J0002.1$+$6721 from the model. Figure~\ref{fig:ts-map} shows the TS map centered at position of 4FGL J0002.1$+$6721. We performed a binned likelihood analysis to obtain the value log(L) and the Akaike information criterion (AIC). The AIC is defined by AIC = -2log(L) + 2k, where k is the number of free parameters in the model. We define \st{the AIC for point source model as 0, and} $\Delta$AIC as the difference value between other models and point source model. There are 18 sources in 4FGL catalog within $5^{\circ}$ of the center position. As we mentioned before, their spectral parameters are set free in our model. In addition, the normalization of Galactic diffuse emission and the isotropic extragalactic emissions are set free also. 

We initiate our analysis by searching \g\ radius of Berkeley 59. A source extension analysis is executed for each source by performing a likelihood ratio test with respect to the point source with a best-fit model for extension: 

\begin{equation}\label{eq:ext}
    TS_{\rm ext} = -2 \left( \mathrm{ln}\left(L_{\rm PS}\right) -  \mathrm{ln}\left(L_{\rm ext}\right) \right)
\end{equation}

For obtaining TS$_{\rm ext}$, we adopt radial Gaussian and radial disk to model the spatial extension of Berkeley 59. With the extension fitting tool \texttt{extension()} from Fermipy to search the radius (95\% confidence interval), we find $1.02 \pm 0.17 ^{\circ}$ and $0.73 \pm 0.11 ^{\circ}$ for 2D radial Gaussian and 2D radial disk, respectively. The TS$_{\rm ext}$ for these two models are 112 and 97 (Table~\ref{tab:fit-model}). 

To investigate whether extended \g\ emission correlates with the gas distribution, we considered a spatial template combining molecular hydrogen (H$_2$), neutral hydrogen (HI) and ionized hydrogen (HII). Detail of gas analysis can be found in Section \ref{sec:gas}. For the H$_2$ + HI + HII template, TS$_{\rm ext}$ = 47 is obtained.

There are two suspected sources appear in the field inside 4FGL J0002.1$+$6721, which are named as Src1 and Src2. We check the possibility of two sources with \texttt{gtfindsrc} and find their coordinates are (0.609$^{\circ}$, 67.309$^{\circ}$) and (1.233$^{\circ}$, 66.894$^{\circ}$), respectively. Then we performed extension check as illustrated in Equation~\ref{eq:ext} and obtained $0.11^{\circ}$ and $0.13^{\circ}$ for Src1 and Src2, respectively. However, considering the TS$_{\rm ext}$ for Src1 (TS$_{\rm ext}$ = 15.4) and Src2 (TS$_{\rm ext}$ = 13.2), and $\Delta$AIC, a two sources model is disfavored. 

\begin{figure}
    \centering
    \includegraphics[width=0.95\linewidth]{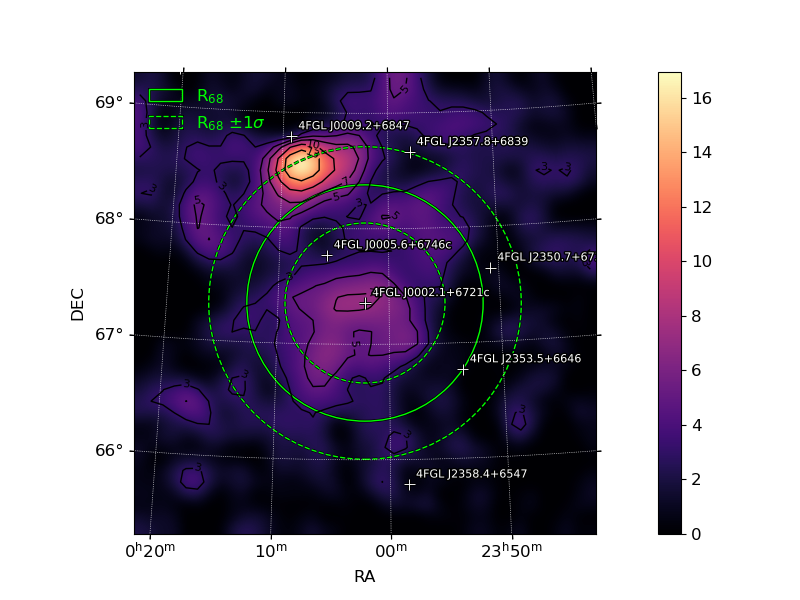}
    \caption{The image of extension check. All the 4FGL sources are marked as cross. The circles represent the estimated radius and the error.}
    \label{fig:ext_check}
\end{figure}

We note that a faint source 4FGL J0005.6$+$6746c (0.5$^{\circ}$ away from 4FGL J0002.1$+$6721) locates inside the radius of 4FGL J0002.1$+$6721 (Fig. \ref{fig:ext_check}). It is necessary to ask whether the \g\ emissions from 4FGL J0005.6$+$6746c is part of Berkeley 59. Therefore, by removing 4FGL J0005.6$+$6746c, we re-examined the extensions and obtained values of $0.85 \pm 0.09^{\circ}$ and  $0.67 \pm 0.05 ^{\circ}$, respectively. Nevertheless, the $\Delta$AIC values of Disk and Gaussian model show that no improvement is obtained for the statistics.

\begin{table*}
    \centering
    \begin{tabular}{l|c|c|c|c|c}
    \hline
        Model & -log(L) & TS$_{\rm ext}$ & d.o.f. & AIC & $\Delta$AIC \\\hline
        Point Source & -3225455 & - & 46 & -6450818 & - \\
        Gaussian & -3225596 & 112 & 47 & -6451098 & -280\\
        Disk & -3225501 & 97 & 47 & -6450908 & -90 \\
        Gaussian (remove 4FGL J0005.6+6746c) & -3225471 & 118 & 45 & -6450852 & -34\\
        Disk (remove 4FGL J0005.6+6746c) & -3225468 & 102 & 45 & -6450846 & -28\\
        H$_2$ + HI + HII & -3225474 & 47 & 50 & -6450848 & -30\\
        0.11$^{\circ}$ Gaussian + 0.13$^{\circ}$ Gaussian & -3225472 & 15.4 (Src1), 13.2 (Src2) & 51 & -6450842 & -24\\
    \hline
    \end{tabular}
    \caption{Analysis results of different spatial models.}
    \label{tab:fit-model}
\end{table*}

\subsection{Spectral Analysis}

The spectral shape aids us to perform a judgment on the nature of \g\ sources. To validate the spectral model, we test for spectral curvature that reveals deviations from a power law (PL) spectrum for each source via likelihood ratio test. For the LogParabola (LP) model, it provides: TS$_{\rm LP}$ = $-2 \left( \mathrm{ln}\left(L_{\rm PL}\right) - \mathrm{ln}\left(L_{\rm LP}\right) \right)$. The PL model is given by: 

\begin{equation}
    \frac{dN}{dE} = N_0 \left( \frac{E}{E_0} \right)^{-\gamma}
\end{equation}
where $N_0$, $E_0$ and $\gamma$ are normalization, scale and photon index, respectively.

The LP model in \lat\ data analysis is defined by:

\begin{equation}
    \frac{dN}{dE} = N_0 \left( \frac{E}{E_b} \right)^{-(\alpha + \beta\ln(E/E_b))}
\end{equation}

where $N_0$ is the normalization, $\alpha$ is the photon index, $\beta$ is the curvature index, and $E_b$ is the reference energy. Similarly, for PLSuperExpCutoff (PLSC) model, the curvature check gives: $
TS_{\rm PLSC} = -2 \left( \mathrm{ln}\left(L_{\rm PL}\right) -  \mathrm{ln}\left(L_{\rm PLSC}\right) \right)$
Here, the PLSC model is defined by: 

\begin{equation}
    \frac{dN}{dE} = N_0 \left( \frac{E}{E_0} \right)^{\gamma_0-\frac{d}{2}\ln\frac{E}{E_0} - \frac{db}{6} \ln^2\frac{E}{E_0} - \frac{db^2}{24} \ln^3\frac{E}{E_0}}
\end{equation}

If either TS$_{\rm LP}$ or TS$_{\rm PLSC}$ exceeds 25, we perform a spectral fit for LP and PLSC models; otherwise, the PL model is adopted. The curvature tests indicate that TS$_{\rm LP} = 0$ and TS$_{\rm PLSC} = 1.6$. Therefore, we use the PL as spectral model to fit the data of Berkeley 59. We note that the 4FGL catalog adopts an LP model for 4FGL~J0002.1$+$6721, whereas our analysis favors a simple PL model. The spectral parameters of the sources within $5^{\circ}$ of 4FGL~J0002.1$+$6721, along with the Galactic and isotropic diffuse emission components are
all set free. We obtain a spectral index $\Gamma = 2.88 \pm 0.06$ and a \g\ luminosity $L_{\gamma} = 1.15 \times 10^{33}\ \rm erg/s$.

We derived the spectral energy distribution (SED), which are shown in Figure~\ref{fig:sed}. We derived the energy range 300 MeV to 300 GeV into six logarithmically spaced energy bins, and in each of them the SED flux is estimated via maximum likelihood analysis. When the TS value of spectral point is less than 9, the upper limit is calculated at the 95\% confidence level. By prforming \textit{gtlike} to finish the maximum likelihood analysis for each energy bins, the index was fixed to 2. The derived \g\ spectrum is consistent with a PL with index $2.88 \pm 0.06$, which is typical for hadronic emission from star-forming regions \citep{Ackermann2011,Sun2020}. The extension of the emission, modeled as a radial Gaussian of $1.02^{\circ}$, suggests that the CRs are diffusing from the cluster and interacting with the surrounding gas. This provides strong evidence for Berkeley 59 being a source of CRs.

\section{Gas Content} \label{sec:gas}

To study the interaction between stellar winds and the surrounding gas, we investigate distribution of molecular hydrogen H$_2$, neutral atomic hydrogen HI, and ionized hydrogen HII in the vicinity of Berkeley 59. 

\begin{figure*}
    \centering
    \includegraphics[width=0.33\textwidth]{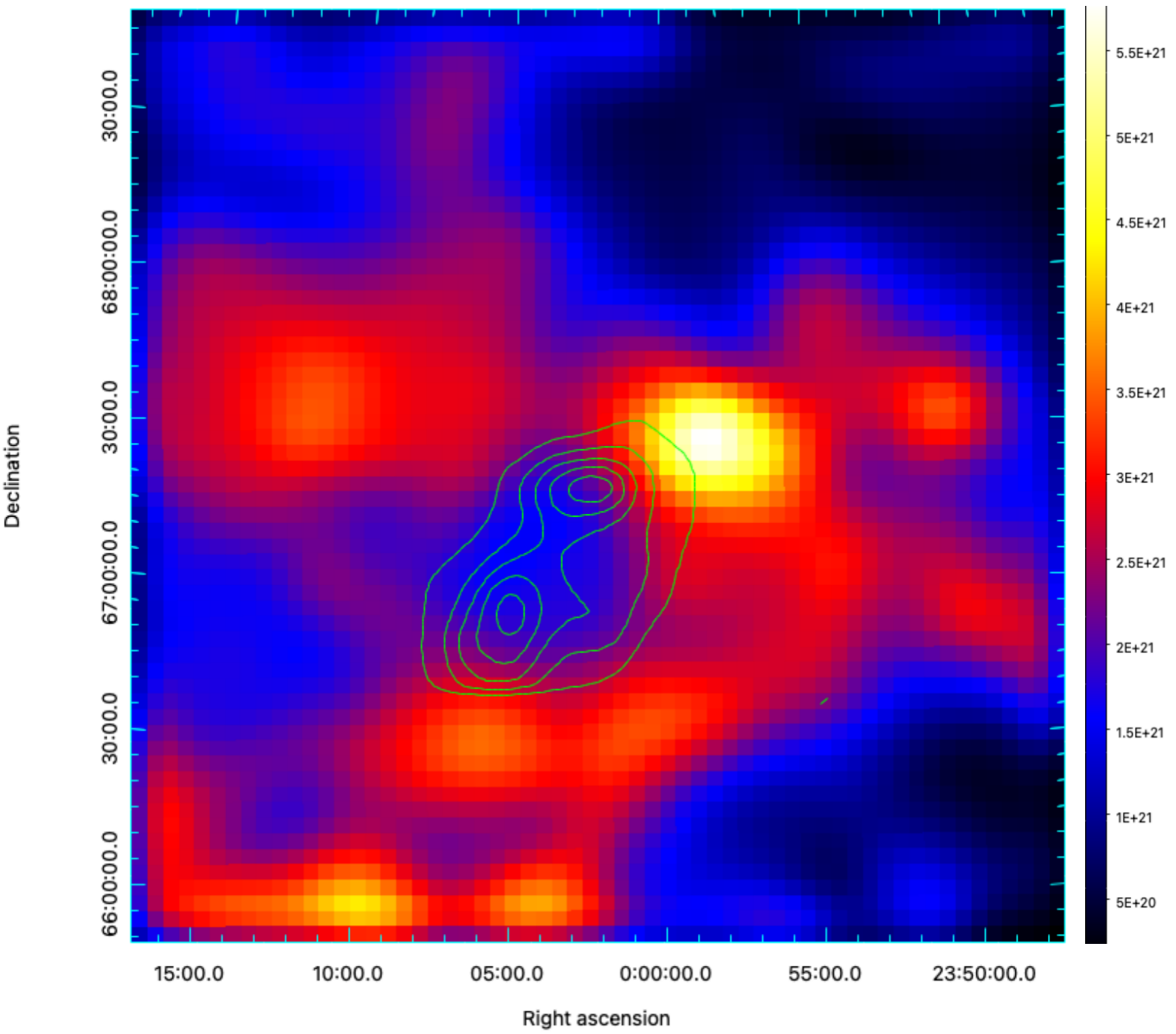}
    \includegraphics[width=0.33\textwidth]{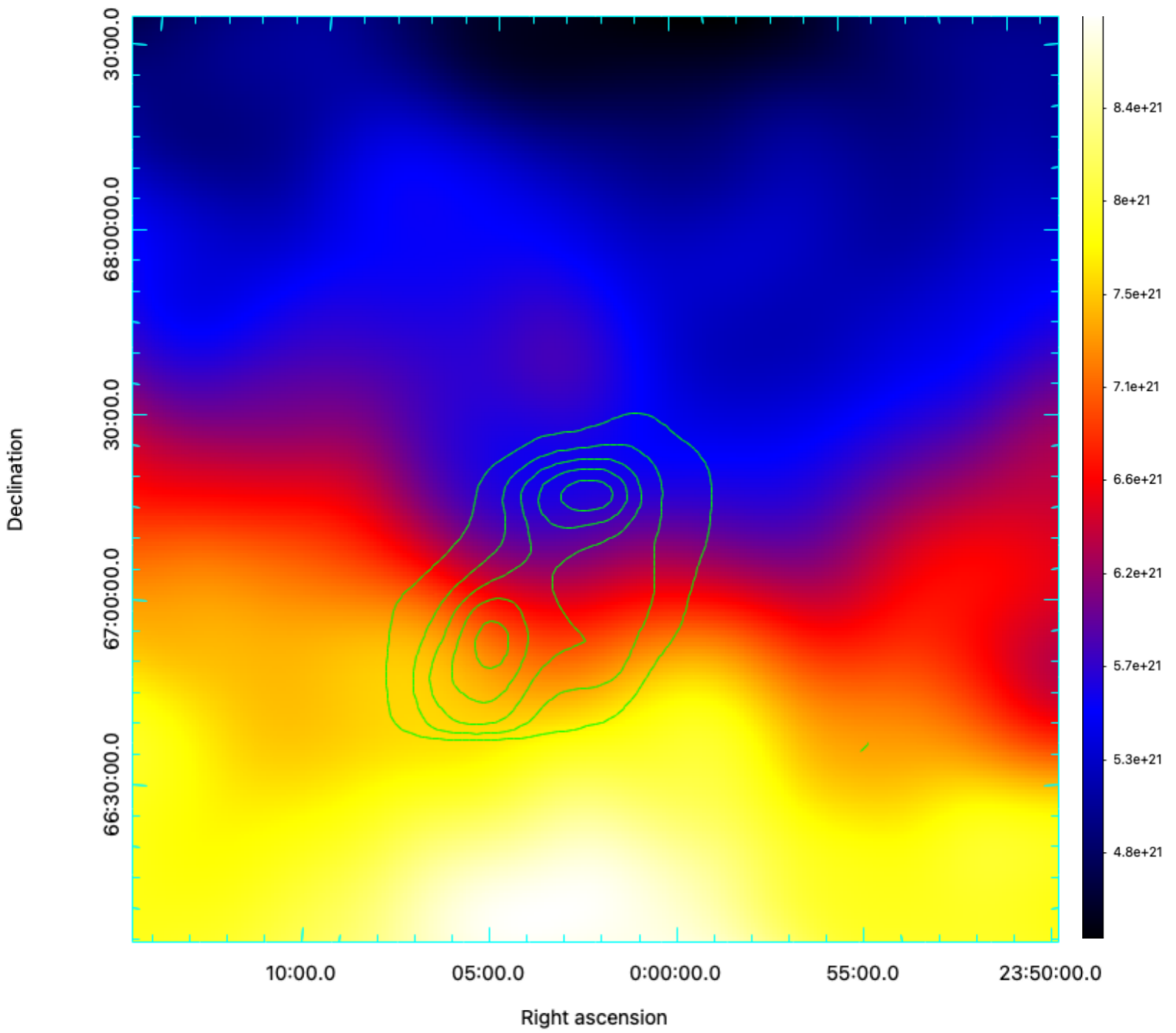}
    \includegraphics[width=0.33\textwidth]{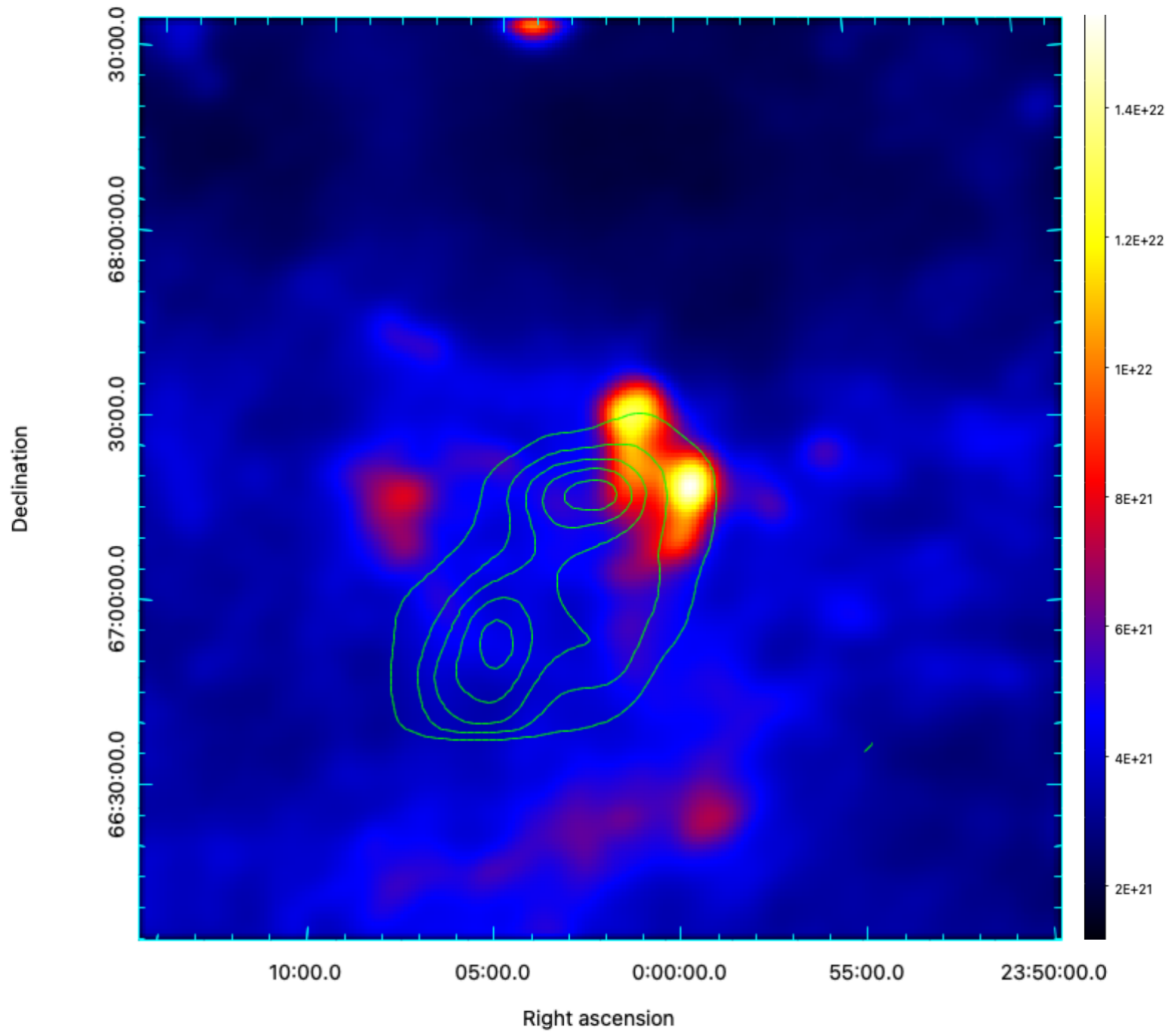}
    \caption{Gas column densities (in unit of $\rm cm^{-2}$) in different gas phase. The left panel gives the $\rm H_2$ column density derived from CO data. The middle panel provides the map of H I column density obtained from a 21-cm all-sky survey. The right panel shows the H II column density derived from \textit{Planck} free-free (353 GHz) map assuming an effective density of electron $n_e = 10\ \rm cm^{-3}$. The \g\ emission is presented as green contours.}
    \label{fig:gas}
\end{figure*}

To investigate the H$_2$ distribution toward Berkeley 59, we use CO data from \cite{Dame2001}. The $^{13}$CO (J = 1-0) line profile of the molecular cloud may reflect the kinematic activity of the gas distribution. For Berkeley 59, the velocity range of the CO is considered to be 4-12 $\rm km\ s^{-1}$ \citep{Gahm2022}. The column density of H$_2$ can be defined by $N_{\rm H_2} = X_{\rm CO} \times W_{\rm CO}$. The conversion factor $X_{\rm CO}$ is set to $2.0 \times 10^{20}\ \rm cm^{-2}\ K^{-1}\ km^{-1}\ s$ \citep{Dame2001,Bolatto2013}. The CO line intensity $W_{\rm CO}$ (in unit of $\rm K\ km\ s^{-1}$) was calculated over the velocity range specified above. Figure \ref{fig:gas} (left) shows the column density of H$_2$.

The HI data are obtained from the HI 4$\pi$ survey (HI4PI) \citep{HI4PI2016}. This survey aims to obtain all-sky Galactic HI data with 21 cm hydrogen lines. The number density $N_{\rm HI}$ can be calculated as:

\begin{equation}
   N_{\rm HI} = -1.38 \times 10^{18}\ T_{\rm s} \int d\nu \ln (1- \frac{T_{\rm B}}{T_{\rm s} - T_{\rm bg}})
\end{equation}

where $T_{\rm B}$ and $T_{\rm bg}$ are brightness temperatures of the HI emission, and the cosmic microwave background, respectively. We adopted $T_{\rm bg} = 2.66\ \rm K$ for the calculation. For $T_{\rm B} > T_{\rm s} - 5\ \rm K$, $T_{\rm B}$ can be approximated as $T_{\rm s} - 5\ \rm K$, and $T_{\rm s}$ is taken to be 150 K. The column density map is presented in Figure~\ref{fig:gas} (middle).

The H II column density can be derived using the \textit{Planck} free-free map \citep{Planck2016}. To do this, the emission measure must be converted into free–free intensity. The conversion factor for this transformation is provided in \cite{Finkbeiner2003}. The column density can then be calculated from the free-free emission: 

\begin{equation} 
    N_{\rm H II}  = 1.2 \times 10^{15} \ (\frac{T_{\rm e}}{1\ \rm K})^{0.35} (\frac{\nu}{1 \ \rm GHz})^{0.1} (\frac{n_{\rm e}}{1\ \rm  cm^{-3}})^{-1} (\frac{I_{\nu}}{1\ \rm Jy\ sr^{-1}})
\end{equation}

$N_{\rm H II}$ is in unit of $\rm cm^{-2}$. In such case, the frequency $\nu = 353\ \rm GHz$, electron temperature $T_{\rm e} = 8000\ \rm K$, and the effective electron density $n_{\rm e} = 10\ \rm cm^{-3}$ are adopted in the calculation. $I_{\nu} = 46.04\ \rm Jy\ sr^{-1}$ is adopted corresponding to $\nu = 353\ \rm GHz$ \citep{Finkbeiner2003}. Obviously, the H II column density is inversely proportional to the $n_{\rm e}$. Figure \ref{fig:gas} (right) gives the derived HII column density.

\begin{table}
    \centering
    \begin{tabular}{l|c|c}
    \hline
        Tracer & Mass ($10^2\ \rm M_{\odot}$) & Number density ($\rm cm^{-3}$)\\\hline
        H$_2$ & 141.2 & 24.18 \\
        H I   & 329.1 & 56.37 \\
        H II  & 164.4 & 28.16 \\
        total & 634.7 & 108.7 \\
        \hline
    \end{tabular}
    \caption{Mass and number density \st{of gas surrounding Berkeley 59} within the radial Gaussian with a radius of $1.02^\circ$.}
    \label{tab:gas}
\end{table}

The total mass of the cloud in each pixel can be calculated with 
\begin{equation}
    M = m_{\rm H} N_{\rm H} A d^2 
\end{equation}
where $m_{\rm H}$ is the mass of the H, $N_{\rm H}$ is the column density in different gas phase, $A$ is the angular area, and $d$ is the distance. Accordingly, the mass \st{of gas surrounding Berkeley 59} is $6.34\times 10^4\ \rm M_{\odot}$ and the average number density is $108.7\ \rm cm^{-3}$ (Table~\ref{tab:gas}). This provides a sufficient target mass for \g\ production via pion decay. Assuming a spherical geometry, the derived mass of Berkeley 59 is $\sim 5.4 \times 10^4 M_{\odot}$.

\section{Cosmic-ray Injection} \label{sec:injection}

The hadronic origin has been discussed in for several YSMCs \citep{Yang2018,Sun2020,Liu2022,Ge2022,Liu2024}. Such \g\ emission may originate from the decay of neutral pions produced by the interactions between the accelerated hadrons and the surrounding gas. Therefore, we explore the scenario in which the \g\ emission is produced via the pion-decay process from the interaction of the CRs with the ambient gas. As provided by Section \ref{sec:gas}, the average number density of the target protons for this region is $108.7\ \rm cm^{-3}$.

\begin{figure}
    \centering
    \includegraphics[width=0.48\textwidth]{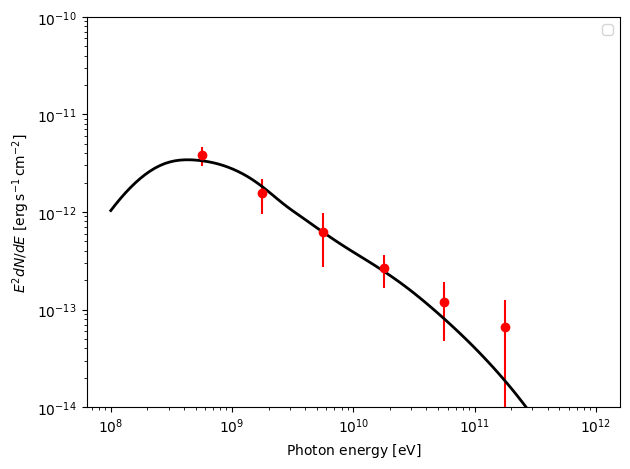}
    \caption{\g\ spectrum of Berkeley 59 with \lat\ data. The line shows the hadronic model.}
    \label{fig:sed}
\end{figure}

If Berkeley 59 can accelerate protons to high energies, then these protons could illuminate the molecular clouds around them via proton–proton interaction. We assume an exponential cutoff power-law spectrum for the parent proton distribution: $f_{\rm p}(E) = A_{\rm p} (E/E_0)^{-\alpha_{\rm p}}\exp (- (E / E_{\rm c})$. We use \texttt{NAIMA} package to fit the SED \citep{Zabalza2015}. Considering the distance of about 1 kpc, the best-fit proton index $\alpha_{\rm p} = 2.91 \pm 0.05$ and the total proton energy $W_{\rm p} = 1.51 \times 10^{49}\ \rm erg$ above 1 GeV are obtained. The model with best-fit parameters are shown in Figure \ref{fig:sed}. 

We also tested the potential leptonic scenario of Berkeley 59. It reflects the \g\ generated via the inverse Compton (IC) scattering of electrons off the seed photons around the source. We considered the CMB radiation field, optical to UV radiation field from the star light as photon field, and the dust infrared radiation field \cite{Kafexhiu2014}. In addition, nonthermal bremsstrahlung radiation resulting from the interaction between relativistic particles and thermal particle population may contribute \g\ emissions \citep{Baring1999}. Accordingly, we calculated the IC and bremsstrahlung spectrum using \texttt{NAIMA} as we did for hadronic scenario. We assumed a exponential cutoff power-law distribution of the relativistic electrons: $f_{\rm e}(E) = A_{\rm e} (E/E_0)^{-\alpha_{\rm e}}\exp (- (E / E_{\rm c})$. The best-fit electron index of electron distribution is $\alpha_{\rm e} = 2.19 \pm 0.04$. We obtain the total energy above 1 GeV: $W_{\rm e} = 2.24 \times 10^{49}\ \rm erg$. We emphasize that the leptonic origin of Berkeley 59 can not be ruled out.

The total kinetic energy injection rate from the wind of the whole stellar cluster is defined by:

\begin{equation}
    L_{\rm w} = (1/2) \sum^{N_{\rm OB}}_{i}\dot{M}_{\rm w} v_{\rm w,~i}^2
\end{equation}

where $N_{\rm OB}$ is number of OB stars, $\dot{M}_{\rm w}$ is the mass-loss rate, and $v_{\rm w,~i}$ is the wind velocity of each OB star. We assume ${\dot{M}_{\rm w}} = 10^{-4} \ M_\odot \ \rm yr^{-1}$ and $v_{\rm w} = 1000 \ \rm km \ s^{-1}$. Then we can derive the kinetic energy injection rate $L_{\rm w} = 9.62 \times 10^{35}\ \rm erg\ s^{-1}$, which is slightly smaller than that derived by \cite{Peron2025a} ($L_{\rm w} = 3.7 \times 10^{36}\ \rm erg\ s^{-1}$). Berkeley 59 has a total stellar mass of $1.0 \times 10^3 \ M_{\odot}$ \citep{Panwar2018}. The energy injection rate into CRs by stellar winds is given by

\begin{equation}
    \dot{E}_{\rm CR} = \eta_{\rm CR} L_{\rm w}
\end{equation}

where $\eta_{\rm CR}$ reflects the fraction of the wind kinetic energy which goes into accelerating CR protons. Here, the $\eta_{\rm CR}$ can be expressed as:
\begin{eqnarray}
    \eta_{\rm CR}\,&\simeq&\,0.4\left(\frac{10^3\, {\rm cm^{-3}}}{n_{\rm eff}}\right)\left(\frac{7\,{\rm pc}}{R}\right)^2 \nonumber \\
    &\,&\times \left(\frac{D}{10^{28}\,{\rm cm^2\,\,s^{-1}}}\right)\left(\frac{3L_\gamma/ L_{\rm w}}{0.01}\right)
\label{eq:high_eta}
\end{eqnarray}

where $n_{\rm eff}$ is the effective number density, $D$ is the diffusion coefficient. Then, the diffusion coefficient $D$ can be written as \citep{Panwar2018}: 

\begin{eqnarray}
    D\,&\simeq&\,2.5\times10^{27}\,{\rm cm^2\,\,s^{-1}} \left(\frac{\eta_{\rm CR}}{0.1}\right)
    \left(\frac{n_{\rm eff}}{10^3\, {\rm cm^{-3}}}\right) \nonumber \\
    &\,&\times\left(\frac{R}{7\,{\rm pc}}\right)^2 
    \left(\frac{0.01}{3L_\gamma/L_{\rm w}}\right)
    \label{bigD}
\end{eqnarray}

Assuming a spherical volume with radius $R = 17.5\ \rm pc$ and total mass of $5.4 \times 10^4 M_{\odot}$, one can obtain estimate $n_{\rm eff} \approx 19\ \rm cm^{-3} $. We use the derived $L_{\gamma} = 1.15 \times 10^{33}\ \rm erg\ s^{-1}$ and $L_{\rm w} = 9.62 \times 10^{35}\ \rm erg\ s^{-1}$. For $\eta_{\rm CR}$, we assume $\eta_{\rm CR} = 0.1$ for typical value of supernova shock and $\eta_{\rm CR} = 0.4$ for high acceleration efficiency, following \cite{Vink2012} and \cite{Pandey2024}. Notice that such assumption has uncertainty, which need to be tested in future observations. The estimated $D$ should be $8.3 \times 10^{26}\ \rm cm^2\ s^{-1}$ and $3.3 \times 10^{27}\ \rm cm^2\ s^{-1}$, respectively. Thus, for the diffusion radius $R_{\rm diff} = \sqrt{D \tau}$ with assumption that the age $\tau$ approximates to the diffusion time scale, we got 148 pc < $R_{\rm diff}$ < 257 pc.

The estimation of the CR injection efficiency and the diffusion coefficient in Berkeley 59 indicates that the stellar winds from the OB stars can provide the necessary energy for CR acceleration. The derived diffusion coefficient is on the order of $10^{27}\ \rm  cm^2\ s^{-1}$, which is slower than the Galactic average, suggesting that the CRs are confined better in the cluster environment. This enhanced confinement could be due to the turbulent magnetic fields in the cluster, which facilitates the acceleration of CRs to high energies. However, the uncertainties in the effective gas density and the acceleration efficiency require further observations to refine.

Considering the kinetic energy injection rate $L_{\rm w} = 9.62 \times 10^{35}\ \rm erg\ s^{-1}$ and the age should be $\tau \approx 2\ \rm Myr$ for Berkeley 59, the total energy injected is obtained as $6.07 \times 10^{49}\ \rm erg$. Thus, this YSMC is powerful enough to accelerate CRs to account for the \g\ emissions detected by \lat\ .

\section{Discussion and Conclusions} \label{sec:conclu}

The estimated CR acceleration efficiency of $\eta_{\rm CR} = 0.1-0.4$ and diffusion coefficient $8.3 \times 10^{26}\ \rm cm^2 \ s^{-1}$ to $3.3 \times 10^{27}\ \rm cm^2 \ s^{-1}$ place Berkeley 59 in the context of other \g\ emitting stellar clusters. This diffusion coefficient is significantly lower than the Galactic average ($\sim 10^{28}\ \rm cm^2\ s^{-1}$), suggesting enhanced CR confinement within the cluster environment, possibly due to turbulent magnetic fields amplified by stellar winds and supernova activity. Such conditions are conducive to efficient particle acceleration and \g\ production.

We report the extended \g\ emissions around the stellar cluster Berkeley 59 using 16 yr of \lat\ data. Insights into the role of stellar clusters in the acceleration and propagation of CRs within our Galaxy is provided. The \g\ source 4FGL J0002.1$+$6721 is potentially associated with Berkeley 59. The extended \g\ emissions has an angular extension of $1.02^{\circ}$ which can be modeled by a radial Gaussian profile.

The total mass of the gas was found to be approximately $6.34 \times 10^4\ M_{\odot}$, with an average target proton density $108.7\ \rm cm^{-3}$. According to our \lat\ analysis, the estimated mass of Berkeley 59 is $\sim 5.4 \times 10^4 M_{\odot}$. The observed bolometric luminosity of the OB associations in these clusters was used to calculate the stellar wind luminosity, $L_{\rm w} = 9.62 \times 10^{35}\ \rm erg\ s^{-1}$. We constraint the CR acceleration efficiency $\eta_{\rm CR}$ and the diffusion coefficient $D$. However, the uncertainty of $\eta_{\rm CR}$ and $D$ need to be refined through a more detailed mapping of the gas distribution in the region. The total energy of injection is at the level of $\sim 10^{49}\ \rm erg$.

\section*{Data Availability}

The \lat\ data used in this work are publicly available and are provided online at the NASA-GSFC Fermi Science Support Center \footnote{https://fermi.gsfc.nasa.gov/cgi-bin/ssc/LAT/LATDataQuery.cgi}. We made use of the CO data to derive the H$_2$ column density \footnote{https://lambda.gsfc.nasa.gov/product/}. The data from Planck Legacy Archive were used to derive the H II column density \footnote{http://pla.esac.esa.int/pla/\#home}. The H I data were taken from the HI4PI \footnote{http://cdsarc.u-strasbg.fr/viz-bin/qcat?J/A+A/594/A116}. They can be found in SkyView also\footnote{https://skyview.gsfc.nasa.gov/current/cgi/query.pl} also. MWISP data could be appied from Millimeter Wave Radio Astronomy Database \footnote{http://www.radioast.cn}, which is operated by the Purple Mountain Observatory, Chinese Academy of Sciences.

\section*{Acknowledgements}

We thank Yuan Li for discussions on molecular gas. We thank Alison Mitchell for providing her comments on stellar clusters. Ziwei Ou is supported by the National Natural Science Foundation of China (NSFC, Grant No. 12393853). Xiaolong Yang is supported by the NSFC (Grant No. 12103076 and 12303054), Shanghai Sailing Program (Grant No. 21YF1455300), and the National Key Research and Development (R\&D) Program of China (Grant No. 2024YFA1611603). Songpeng Pei is supported by the  Science and Technology Foundation of Guizhou Province (Grant No. QKHJCMS[2026]752) and the Liupanshui Science and Technology Development Project (Grant No. 52020-2025-0-2-13)..



\bibliographystyle{mnras}
\bibliography{example} 


\bsp	
\label{lastpage}
\end{document}